\documentclass[useAMS,usenatbib]{mn2e}
\usepackage{amsmath}
\usepackage{amssymb}
\usepackage{amsfonts}
\usepackage{xcolor}
\usepackage[draft]{graphicx}

\bibpunct{(}{)}{;}{a}{,}{,}

\makeatletter
\def\ref@jnl#1{{\rmfamily#1}}%
\newcommand\aj{\ref@jnl{AJ}}%
\newcommand\araa{\ref@jnl{ARA\&A}}%
\newcommand\apj{\ref@jnl{ApJ}}%
\newcommand\apjl{\ref@jnl{ApJ}}%
\newcommand\apjs{\ref@jnl{ApJS}}%
\newcommand\ao{\ref@jnl{Appl.~Opt.}}%
\newcommand\apss{\ref@jnl{Ap\&SS}}%
\newcommand\aap{\ref@jnl{A\&A}}%
\newcommand\aapr{\ref@jnl{A\&A~Rev.}}%
\newcommand\aaps{\ref@jnl{A\&AS}}%
\newcommand\azh{\ref@jnl{AZh}}%
\newcommand\baas{\ref@jnl{BAAS}}%
\newcommand\jrasc{\ref@jnl{JRASC}}%
\newcommand\memras{\ref@jnl{MmRAS}}%
\newcommand\mnras{\ref@jnl{MNRAS}}%
\newcommand\pra{\ref@jnl{Phys.~Rev.~A}}%
\newcommand\prb{\ref@jnl{Phys.~Rev.~B}}%
\newcommand\prc{\ref@jnl{Phys.~Rev.~C}}%
\newcommand\prd{\ref@jnl{Phys.~Rev.~D}}%
\newcommand\pre{\ref@jnl{Phys.~Rev.~E}}%
\newcommand\prl{\ref@jnl{Phys.~Rev.~Lett.}}%
\newcommand\pasp{\ref@jnl{PASP}}%
\newcommand\pasj{\ref@jnl{PASJ}}%
\newcommand\qjras{\ref@jnl{QJRAS}}%
\newcommand\skytel{\ref@jnl{S\&T}}%
\newcommand\solphys{\ref@jnl{Sol.~Phys.}}%
\newcommand\sovast{\ref@jnl{Soviet~Ast.}}%
\newcommand\ssr{\ref@jnl{Space~Sci.~Rev.}}%
\newcommand\zap{\ref@jnl{ZAp}}%
\newcommand\nat{\ref@jnl{Nature}}%
\newcommand\iaucirc{\ref@jnl{IAU~Circ.}}%
\newcommand\aplett{\ref@jnl{Astrophys.~Lett.}}%
\newcommand\apspr{\ref@jnl{Astrophys.~Space~Phys.~Res.}}%
\newcommand\bain{\ref@jnl{Bull.~Astron.~Inst.~Netherlands}}%
\newcommand\fcp{\ref@jnl{Fund.~Cosmic~Phys.}}%
\newcommand\gca{\ref@jnl{Geochim.~Cosmochim.~Acta}}%
\newcommand\grl{\ref@jnl{Geophys.~Res.~Lett.}}%
\newcommand\jcp{\ref@jnl{J.~Chem.~Phys.}}%
\newcommand\jgr{\ref@jnl{J.~Geophys.~Res.}}%
\newcommand\jqsrt{\ref@jnl{J.~Quant.~Spec.~Radiat.~Transf.}}%
\newcommand\memsai{\ref@jnl{Mem.~Soc.~Astron.~Italiana}}%
\newcommand\nphysa{\ref@jnl{Nucl.~Phys.~A}}%
\newcommand\physrep{\ref@jnl{Phys.~Rep.}}%
\newcommand\physscr{\ref@jnl{Phys.~Scr}}%
\newcommand\planss{\ref@jnl{Planet.~Space~Sci.}}%
\newcommand\procspie{\ref@jnl{Proc.~SPIE}}%

\long\def\comment#1{}

\def\ba{\begin{eqnarray}}
\def\ea{\end{eqnarray}}
\def\be{\begin{equation}}
\def\ee{\end{equation}}

\def\mA{\mathbfss{A}}
\def\mB{\mathbfss{B}}

\def\mD{\mathbfss{D}}
\def\mF{\mathbfss{F}}
\def\mG{\mathbfss{G}}
\def\mI{\mathbfss{I}}
\def\mS{\mathbfss{S}}
\def\mM{\mathbfss{M}}

\def\mO{\mathbfss{O}}
\def\mP{\mathbfss{P}}
\def\mR{\mathbfss{R}}
\def\mT{\mathbfss{T}}
\def\mU{\mathbfss{U}}
\def\mV{\mathbfss{V}}
\def\mW{\mathbfss{W}}
\def\mL{\mathbfss{L}}
\def\mLambda{\mathbf{\Lambda}}

\def\inv{^{-1}}

\def\bdx{\bmath{x }}

\def\bdv{\bmath{v }}
\def\bda{\bmath{a }}
\def\bdb{\bmath{b }}

\def\bdn{\bmath{n }}

\def\bdf{\bmath{f }}
\def\bdg{\bmath{g }}
\def\bdb{\bmath{b }}
\def\bdN{\bmath{N }}
\def\bdy{\bmath{y }}

\def\veca{{\bda}}
\def\vecb{{\bdb}}

\def\vecf{{\bdf}}
\def\vecg{{\bdg}}

\newcommand{\tR}{{\mR}}
\newcommand{\tA}{{\mA}}
\newcommand{\tF}{{\mF}}
\newcommand{\tG}{{\mG}}
\newcommand{\tT}{{\mT}}
\newcommand{\tO}{{\mO}}
\newcommand{\tD}{{\mD}}
\newcommand{\tP}{{\mP}}
\newcommand{\tU}{{\mU}}
\newcommand{\tV}{{\mV}}

\newcommand{\tB}{{\mB}}

\newcommand{\tW}{{\mW}}
\newcommand{\tI}{{\mI}}
\newcommand{\tS}{{\mS}}
\newcommand{\tM}{{\mM}}
\newcommand{\tL}{{\mL}}
\newcommand{\tLambda}{{\mLambda}}

\newcommand{\tRfg}{\mR_{\mF\mG}}

\def\nobs{N_{\rm obs}}

\def\rfg{N_{\rm FG}}

\def\Col{\mathrm{Col}}
\def\Span{\mathrm{Span}}

\title[Foreground component separation with generalised ILC]{Foreground component separation with generalised ILC}
\author[Mathieu Remazeilles, Jacques Delabrouille, Jean-Fran\c{c}ois Cardoso]{Mathieu Remazeilles\thanks{E-mail: remazeil@apc.univ-paris7.fr}, Jacques Delabrouille\thanks{E-mail: delabrouille@apc.univ-paris7.fr}, Jean-Fran\c{c}ois Cardoso\thanks{E-mail: cardoso@enst.fr}\\
APC
10, rue Alice Domon et L\'eonie Duquet,
75205 Paris Cedex 13,
France}

\begin{document}


\pagerange{\pageref{firstpage}--\pageref{lastpage}} \pubyear{2010}

\maketitle

\label{firstpage}

\begin{abstract}
The `Internal Linear Combination' (ILC) component separation method has been extensively used to extract a single component, the CMB, from the WMAP multi-frequency data.
We generalise the ILC approach for separating other millimetre astrophysical emissions. We construct in particular a multidimensional ILC filter, which can be used, for instance, to estimate the diffuse emission of a complex component originating from multiple correlated emissions, such as the total emission of the Galactic interstellar medium. The performance of such generalised ILC methods, implemented on a needlet frame, is tested on simulations of Planck mission observations, for which we successfully reconstruct a low noise estimate of emission from astrophysical foregrounds with vanishing CMB and SZ contamination. 
\end{abstract}


\section{Introduction}

The separation of emissions originating from distinct astrophysical components in observations of the millimetre and sub-millimetre sky is an important step in the scientific exploitation of such observational data. Various methods have been developed to extract the emission of several components out of multi-frequency Cosmic Microwave Background (CMB) observations such as those of the WMAP and Planck space missions (see, e.g., \citet{2009LNP...665..159D} for a review). 

In many cases, such methods define components through an (explicit or implicit) assumption that the observations are a linear mixture of unknown templates (or \emph{sources}) scaling rigidly with frequency, i.e.:
\begin{equation}
y_i(p)  =  \Sigma_{j} A_{ij} s_j(p) + n_i(p).
\label{eq:linear-mixture}
\end{equation}
Such methods also assume a fixed number of astrophysical emissions (e.g. CMB anisotropies, thermal Sunyaev-Zel'dovich (SZ) effect, thermal dust emission, synchrotron emission...). The rigid scaling of component emission with frequencies is imposed by the fact that the mixing coefficients $A_{ij}$ depend solely on $i$ and $j$ (observation channel and component), and not on the pixel $p$.

Assuming that such a representation holds, blind component separation
methods such as the Spectral Matching ICA
\citep{2003MNRAS.346.1089D,2008ISTSP...2..735C}, FastICA
\citep{FastICA,2002MNRAS.334...53M}, JADE \citep{JADE}, CCA
\citep{2006MNRAS.373..271B} or GMCA \citep{2008StMet...5..307B} are
designed to solve the problem of recovering the components of interest
when their \emph{mixing matrix} (the matrix of mixing coefficients, which specifies how much  each component contributes to a frequency observation) is unknown. 
By exploiting the assumption of statistical independence between the
components, the mixing matrix can be blindly estimated up to permutation
and rescaling of its columns. Once an estimate of the mixing matrix is available, the components can be separated by inverting the linear
system,
possibly taking into account the
presence of instrumental noise.  This has been investigated by a
number of authors \citep{1996MNRAS.281.1297T,
  1999NewA....4..443B,1998MNRAS.300....1H,2002MNRAS.330..807D}.

However, in millimetre and sub-millimetre wave observations, some
components cannot be correctly modelled as a single template which
would be simply scaled by mixing coefficients \citep{1998ApJ...502....1T}.  Emissions from the
Galactic interstellar medium exhibit frequency scaling which depends
on local conditions (temperature, chemical composition) at the
location of emission, and hence are variable over the celestial
sphere.

Some of the blind component separation methods quoted above can take into account the possible variation of the foreground frequency scaling as a function of the observed pixel. The CCA method, for instance,  can use a pixel-localized model of the foreground spectral indices. The Spectral Matching ICA can be (and has been) implemented on wavelet frames. All methods can be applied independently on several regions of the sky, allowing for a different parameter set in each of the selected regions. However, such localisation of the model and of the solution is then the result of prior choices. The number of foreground components is fixed, the regions to be masked, or to be analysed separately, are selected a priori. In reality, the number of relevant components is not well known before the data is analysed, and also varies in practice both over the sky and over the scales. For instance, most of the galactic foreground components become small, and possibly negligible, at small angular scales and at high Galactic latitude. The number of relevant components in any data set, however, is bound to depend on the level of instrumental noise.

The total foreground emission can be separated by subtracting a CMB map estimate from the observation maps. This method has been investigated on the WMAP observations by \cite{ghosh10} and employed so far by the Planck collaboration in their early results \citep{2011arXiv1101.2028P}. However, the CMB-free maps suffer from excessive noise contamination: as the removed CMB map itself is a low noise estimate obtained from a minimum variance procedure, most of the instrumental noise ends-up in the foreground maps, which must then be re-processed (e.g. filtered) after CMB subtraction.
The new component separation method investigated in this paper is an extension of the Internal Linear Combination (ILC) method, aimed at the reconstruction of the total foreground emission with the intention of both relaxing the prior assumption on the number of foreground components, and of performing a local processing for best suppression of the contamination of the reconstructed foreground components by residual instrumental noise.

Classical ILC methods do not assume a particular
parametrisation of the foreground emission.  They offer a simple way to
extract the map of a single component of interest and have been used
by several authors in the analysis of the maps obtained by the WMAP
satellite to extract a CMB map \citep{2003ApJS..148...97B,
  2004ApJ...612..633E, 2007ApJ...660..959P, 2008arXiv0803.1394K,
  2009A&A...493..835D}.  The traditional ILC, however, can only
recover components for which the emission scales rigidly with
frequency (hence its use for separating a CMB map).  In addition, the ILC performs satisfactorily only if the
component of interest is not correlated with the other emissions.

In a previous publication \citep{2011MNRAS.410.2481R}, we have introduced
the \emph{Constrained ILC}, which extends the ILC to the case where there is more than one component
of interest (e.g. CMB and thermal SZ), and one wishes to cancel out
the contamination from one of them into the recovered map of the
other.  In the present paper, we generalise further the ILC and
address the blind separation of multidimensional components which cannot be
modelled as one single template scaling with frequency according to a single (pixel independent) emission law.

\section{CMB estimation by standard ILC}

\subsection{Model of the measurement}

In all of the following, we assume that all available maps ($\nobs$
maps) can be written, for all pixels $p$ of the observed maps, in the
form
\begin{equation}
  \bdy(p) = \bda s(p) +  \bdb z(p) + \bdf(p) + \bdn(p)
  \label{eq:model}
\end{equation}
where $s(p)$ is the CMB template map, $z(p)$ the thermal
Sunyaev-Zel'dovich effect, $\bdf(p)$ is the emission of the rest of
the foregrounds as they would be observed by the instrument in absence
of anything else, and $\bdn(p)$ is the instrumental noise. Note that
$\bdf(p)$ and $\bdn(p)$ are represented with $\nobs$ maps each, while
the CMB and the SZ effect are represented by one single map each,
scaled across frequency channels using CMB and SZ scaling
coefficients, $\bda$ and $\bdb$, which are assumed to be known.

Depending on the objective, any of $\bda s(p)$, $\bdb z(p)$ or
$\bdf(p)$ can be considered as `noise' and implicitly included in the
noise term.  Similarly, depending on the objectives of the component
separation, $\bda s(p)$ or $\bdb z(p)$ can be considered as part of
the total `foreground term' \textit{i.e.} implicitly included
in~$\bdf(p)$.

\subsection{Extraction of the CMB}

The ILC provides the estimate $\hat{s}_{\rm ILC}$ of the CMB component $s$ by forming a linear combination of the $\nobs$ observed maps which has unit response to the component of interest and has minimum variance. Straightforward algebra leads to:
\be
\hat{s}_{\rm ILC}  = \frac{\bda^t \, {\widehat{\tR}}^{-1}}{\bda^t \, {\widehat{\tR}}^{-1} \, \bda} \, \bdy
\label{eq:ILC}
\ee where ${\widehat{\tR}}$ is the empirical covariance matrix of the
observations, $\bda$ has dimension $\nobs \times 1$, and $\bdy$ is the
$\nobs \times 1$ vector of observation maps.  This standard ILC can be
used similarly to recover an estimate $\hat{z}_{\rm ILC}$ of the SZ
effect (with $\bda$ replaced by $\bdb$ in Eq. (\ref{eq:ILC})). Note
that the quality of CMB reconstruction with an ILC depends on the
accuracy with which $\veca$ is known. 
In presence of errors (for instance calibration errors), there is no
guarantee that the CMB is preserved~\citep{2010MNRAS.401.1602D}.

Assuming no correlations between the components, the total covariance matrix $\tR$ of the observations $\bdy$ can be written as:
\begin{equation}
  \tR = \bda \bda^t C_{\rm CMB} + \bdb \bdb^t C_{\rm SZ} + \tRfg + \tR_{\bdN}  
\end{equation}

\subsection{Wavelet space ILC}

In its simplest implementation, the ILC is performed on the complete maps, and one single global matrix ${\widehat{\tR}}$ is used for the whole data set. This requires all maps to be at the same resolution. It is possible, however, to decompose the original maps as sums of different data subsets, covering each a different region in pixel space or in harmonic space, to apply independent versions of the ILC  to the different data subsets, and then to recompose a map from all these independent results. 

The main interest of such a decomposition is the possibility to adapt
the ILC filter to local contamination conditions. Such localisation of
the filter is useful in pixel space: the Galactic emissions are
stronger in the Galactic plane, whereas noise dominates the total
error at high Galactic latitudes. It is also useful in harmonic space,
because contaminants do not all have the same angular power spectrum
and because of the channel-dependence of instrumental beams.

Note however that some care must be taken when subdividing the original data into small subsets. The ILC indeed relies on the component of interest to be uncorrelated with the contaminants (i.e. $\langle s(p) n_i(p) \rangle = 0$, for all channels of observation $i$). If this condition does not hold, the ILC introduces a bias in the reconstruction. This has a consequence on the minimum data size on which the ILC should be implemented: too few independent data points results in empirical correlations between the component of interest and the contaminants, which generates a reconstruction bias as described in the appendix of \cite{2009A&A...493..835D}.

In the present paper, observations are decomposed using the spherical needlets discussed, in the context of CMB data analysis, by several authors (see, e.g., \citet{2006PhRvD..74d3524P,2008MNRAS.383..539M,PhysRevD.78.083013,guilloux:fay:cardoso:2008}). This needlet decomposition provides localisation of the ILC filters both in pixel and in harmonic space.

We define a set of spectral windows $h^{(j)}(\ell)$ such that, over the useful range of $\ell$, we have:
\begin{equation}
\sum_j \left[h^{(j)}_\ell\right]^2 = 1
\end{equation}
Maps of wavelet (needlet) coefficients are obtained, for each observed
map $y(p)$, by inverse spherical harmonic transform (SHT) of the associated map SHT
coefficients $y_{\ell m}$ filtered by the spectral windows
$h^{(j)}_\ell$:
\begin{equation}
  \gamma^{(j)}(p) = \sum_\ell \sum_m y_{\ell m} \, h^{(j)}_\ell \, Y_{\ell m}(p)
\end{equation}
For each scale $j$, for each pixel $p$ of the corresponding needlet
coefficients maps $\gamma_{a}^{(j)}$ (one such map for each
observation $a$), the empirical covariance matrix ${\widehat{\tR}}$
used in equation (\ref{eq:ILC}) is computed from an average, in a
domain $\mathcal{D}_p$ centered at pixel $p$ and including some
neighbouring pixels, of the product of needlets coefficients. The $ab$
entry is given by
\begin{equation}
  {\widehat{\tR}}_{ab}(p) = \frac{1}{N_p} \sum_{p' \in D_p} \gamma_{a}^{(j)}(p')\gamma_{b}^{(j)}(p')
\end{equation}
In \cite{2009A&A...493..835D}, the practical implementation uses, as
domains $\mathcal{D}_p$, HEALPix `super-pixels' obtained by grouping
$32 \times 32$ pixels of the needlet coefficient maps
$\gamma^{(j)}(p)$ (making use of the hierarchical definition of
HEALPix pixels). Here, we use a slightly different prescription: we
smooth instead the product map
$\gamma_{a}^{(j)}(p)\gamma_{b}^{(j)}(p)$ with a symmetric, Gaussian
window in pixel space. This avoids artificial discontinuities at
super-pixel edges.

\section{Foreground estimation by multidimensional ILC}

We now address the problem of estimating the set of maps $\bdf(p)$,
i.e. a `catch-all' foreground component comprising the emission of the
diffuse Galactic interstellar medium (ISM), and of numerous Galactic
and extragalactic compact sources.   The  objective is to construct
estimated maps $\widehat \bdf(p)$ that `best match' what would be
observed by the instrument in the absence of CMB, SZ and noise (see
Eq.~\ref{eq:model}).

Astrophysical emission originating from the Galactic ISM and from
numerous extragalactic sources is qualitatively different from the CMB
and the SZ effect. Each of the latter is somewhat special in the sense
that its emission can be modeled, with good accuracy, as a single
template scaling in a known way with frequency. The total foreground
(FG) emission $\bdf$ comprises contributions from several different
processes. In addition, we cannot even assume \emph{a priori} that a
linear mixture model (in which each map constituting~$\bdf$ would be a
linear superposition of well defined templates) does hold.

For extracting such emissions from multi-frequency observations, we propose to generalize the ILC method to address the case of such a
`multidimensional component'.  

\subsection{Multidimensional components}\label{sec:multidim-comp}

Let $\tRfg = \langle \bdf \bdf^t \rangle$ denote the covariance matrix of the observed foregrounds in $\nobs$ frequency channels. This $\nobs\times \nobs$ matrix $\tRfg$ will be refered to as the \emph{FG covariance matrix}.

Among astrophysical foregrounds included in the `catch-all' component $\bdf$, the ISM of our own galaxy is the main contributor. It emits via the combination of several processes
(synchrotron, free-free, thermal dust, `anomalous' dust emission, molecular
lines...).  In previous work, some of these processes have been
individually modelled each by a fixed template and an emission
law. \citet{1999NewA....4..443B}, for instance, assume that
synchrotron emission scales with frequency proportionally to
$\nu^{-0.9}$, free-free proportionally to $\nu^{-0.16}$, and dust
proportionally to $\nu^2B_\nu(T)$ with $T=18\, {\rm K}$.  
Since the emission of the ISM in each channel is described as a linear
mixture of three templates, such a model predicts that the ISM
covariance matrix (which we will denote as $\tR_{\mI\mS\mM}$) is a rank 3 matrix.
When the contribution of extragalactic compact sources is neglected (assuming bright point sources are extracted from the maps, and faint ones contribute a negligible amount of emission), the foreground covariance matrix itself, $\tRfg$, is equal to $\tR_{\mI\mS\mM}$ (the covariance matrix of the Galactic ISM emission), and is a rank 3 matrix.

Such a model is too crude in the context of the very
sensitive measurements performed by WMAP and Planck: the emission laws of the Galactic emissions vary as a function of the direction on the sky. To make things even more complex, the background of compact sources contributes emission that becomes significant for measurements such as those of the Planck mission \citep{2011arXiv1101.2028P}, and that can not be modelled at all as the sum of a few independent components.

The question of the rank of the FG covariance matrix is a crucial one for component separation. This matrix is expected to be, strictly speaking, full rank. In practice however, the issue is slightly more subtle.
Consider its eigen-decomposition: $\tRfg = \tV \tD
\tV^t$, where $\tV$ is an orthonormal matrix and $\tD$ is a diagonal
matrix with eigen-values sorted in decreasing order.  While the three-component model of \citet{1999NewA....4..443B}, predicts that
only the first three eigen-values are non-zero, a model with
spatially varying spectral indices, and numerous additional emission processes 
(`anomalous' dust emission, molecular lines, extragalactic source background) predicts that all the eigen-values
are non-zero.  However, if there is only a small variation of the
spectral indices, and if some components are very weak, it is expected, at least in some regions of the sky or at some angular scales, that the smallest eigen-values are
very close to zero so that $\tRfg$ is `almost rank-deficient' (see section~\ref{subsec:dim-FG} below for a more rigorous statement).

In this paper we propose, as in~\cite{2008ISTSP...2..735C}, to model
the FG covariance matrix as an $\nobs\times \nobs$ matrix of rank $\rfg$,
not necessarily equal to $\nobs$.
Loosely speaking, $\rfg$ counts the number of different templates
needed to represent \emph{most} of the emission of the FG in our data set.
In other words, we try to capture all foreground emission as resulting
from $\rfg$ (possibly correlated) templates.  The integer $\rfg$ is
called the (effective) FG dimension and may vary over the sky with respect to the pixel $p$, or with respect to $\ell$ in harmonic space, or with respect to the needlets domain considered for a decomposition of the maps on a needlet frame. 

\subsection{The foreground subspace}
\label{sec:ism-subspace}

The analysis in this paper is performed on a needlet frame. The temperature map needlet coefficients are indexed by $(j,k)$, where $j$ denote the scale and $k$ the pixel.\footnote{Note that the methods described throughout the paper do not require a needlet frame in particular and can be implemented in pixel space as well, where domains $\mathcal{D}$ should be indexed by pixels $p$, or in harmonic space with the domains indexed by $(\ell,m)$ coefficients.}

In a given needlet domain $\mathcal{D}^{(j)}_k$,  if the FG covariance matrix $\tRfg$ is a (symmetric, non
negative) $\nobs\times \nobs$ matrix of rank $\rfg$, then foreground
emission can be represented as a superposition of $\rfg$ templates:
\begin{equation}
  \label{qqq:defa}
  \vecf = \tF~ \vecg
\end{equation}
where the $\nobs \times \rfg$ matrix $\tF$ is called the
\emph{foreground mixing matrix} and where $\vecg$ is a vector of
dimension $\rfg$.  It follows that the FG covariance matrix is
\begin{equation}\label{eq:Rfg}
  \tRfg
  = \langle \vecf\vecf^t \rangle 
  = \tF \langle \vecg\vecg^t \rangle \tF^t 
  = \tF \tG \tF^t
\end{equation}
where $\tG$ is a $\rfg\times \rfg$ full rank covariance matrix.

Note two important points.  
First, the templates $\vecg$ are not expected to correspond to physical foregrounds.  They are just a basis of the
$\rfg$-dimensional subspace spanned by~$\bdf$.  We are not interested in recovering $\vecg$. Our objective with the method
discussed here is to recover $\bdf$ (in addition to $s$ and $z$).
Secondly, matrix $\tF$ and its number $\rfg$ of columns may depend on the domain $\mathcal{D}^{(j)}_k$ considered.  For
instance, at high Galactic latitude, it is quite possible that our observations contain negligible emission from some of
the Galactic foregrounds, but not so at low Galactic latitude. The needlet implementation allows us to modulate the
effective dimension of the multidimensional foreground component both in pixel space and in harmonic space, i.e. vary
$\rfg$ across the sky regions and the physical scales.

We also stress that, unlike in the case of CMB and SZ reconstruction, where the mixing vectors $\bda$ and $\bdb$ are
assumed fully known \emph{a priori}, we do not assume here that the matrix $\tF$ is known.  We will not resort to prior
physical knowledge about the components of the FG emission to determine matrix $\tF$. In fact, as the basis templates
$\vecg$ do not correspond to anything physically meaningful, we are not even interested in determining $\tF$ itself but,
for reconstruction purposes, only the product $\bdf=\tF\bdg$.
It is only assumed that matrix $\tRfg$ has a given rank $\rfg$ (which can be estimated from the data, if needed) in the needlet domain.

Matrix $\tF$ cannot be determined from the data only, that is, without making use of some prior information or assumption
about $\vecg$.  Indeed, let $\tT$ be some invertible $\rfg\times \rfg$ matrix and consider the transformed matrices
$\widetilde \tF = \tF \tT^{-1}$ and $\widetilde \tG = \tT \tG \tT^t$.  These transformed matrices are an alternate,
completely equivalent, factorization of the FG covariance matrix since, by construction, $\tF \tG \tF^t = \widetilde \tF
\widetilde \tG \widetilde \tF^t$.
Physically, it means that the $\rfg$ underlying templates $\vecg$ can be replaced by any other linear combination $\tT
\vecg$ of them (provided the linear combination is not degenerate, i.e. $\tT$ is invertible).

However, as we shall see in section~\ref{subsec:multidim-ilc}, the implementation of the ILC filter for estimating the
total FG emission does not require the full knowledge of $\tF$.  Indeed, the expression of that
filter is strictly unchanged upon the introduction of such an invertible factor $\tT$.
In section~\ref{subsec:dim-FG}, we show how matrix $\tF$ can be estimated up to multiplication by a right factor $\tT$.  It
is worth stressing again that this indetermination means that we are only concerned with estimating the \emph{column space} 
of matrix $\tF$ (noted $\Col(\tF)$ throughout the paper). That $\rfg$-dimensional space can be called the `FG subspace'.
Our working assumption that $\tRfg$ has rank $\rfg$ means that the FG data has a covariance structure which is unknown but
is constrained to live in the FG subspace.

Physically, accepting this indetermination amounts to giving up, during the component separation stage discussed here,
distinction between processes of emission on the basis of physical criteria such as emission process or physical origin.
Obviously, this is not fully satisfactory from an astrophysicist's point of view, since in the end we would like to know
what is the source of the observed emission.  This distinction among sources of FG emission, however, can be made at a
later stage of the data analysis, \textit{i.e.} we can first separate CMB and SZ from other foregrounds, and then put
physics into the interpretation of the reconstructed multidimensional FG component and interpret it as the sum of emissions
from a number of physical emission processes.

\subsection{Multidimensional ILC filter}\label{subsec:multidim-ilc}

Aiming at a direct estimation of the foregrounds, we generalize the ILC method to address the case of a multidimensional
component (here $\rfg$--dimensional, where $\rfg$ is the number of components, i.e. the rank of the foreground covariance
matrix). In a given needlet domain, we model the observation maps, collected into the $\nobs\times 1$ vector $\bdy$, as
\begin{equation}
  \label{qqq:modela}
  \bdy = \tA \bdx + \bdn,  
\end{equation}
where $\bdn$ is the $\nobs\times 1$ vector of instrumental noise and
\begin{equation}
  \label{qqq:model}
  \tA = \left[
\begin{array}{cc}
\bda & \tF
\end{array}\right],\qquad 
\bdx=\left[
\begin{array}{c}      
s \\
\bdg 
\end{array}\right].
\end{equation}
Note that no rigid scaling with frequency is assumed since all these quantities are needlet-dependent, i.e. they depend both on the scale considered and on the pixel.
Here the $(\rfg+1)\times 1$ signal vector $\bdx$ contains the CMB emission $s$ as first entry and the $\rfg\times 1$ vector
$\bdg$ which collects the emission of the $\rfg$ components needed to model the total foreground emission.  The
$\nobs\times (\rfg+1)$ mixing matrix $\tA$ contains, as a first column, the $\nobs\times 1$ vector $\bda$ giving the
frequency scaling of the CMB component.  
The other columns correspond to the $\nobs\times
\rfg$ foreground mixing matrix $\tF$, \textit{i.e.} they span the
foreground subspace. Note that this assumes that $\bda$ itself cannot be obtained
by linear combinations of the columns of $\tF$ (more about this later).

As a refinement, it can be useful to single out both the CMB and the SZ, in which case the second column in $\tA$
explicitly appears as the frequency scaling vector $\bdb$ of the SZ component (and the SZ can be considered as excluded
from the rest of the foregrounds). We get back to this refinement in sections~\ref{subsubsec:sz-orth} and
\ref{sec:results}.

Eq.~(\ref{qqq:modela}) assumes that all observations are at the same resolution, which is needed to implement the ILC
filter (for practical implementation, maps are put to the same resolution by partial deconvolution in harmonic space). The
localisation in harmonic space allows dropping out some of the channels at high $\ell$ if needed by reason of insufficient
resolution.

We consider the estimation of $\vecf$ by a linear operation
\begin{equation}
  \label{eq:2}
  \widehat{\vecf} =\tB \bdy,
\end{equation}
where, as in standard (one-dimensional) ILC, the $\nobs\times \nobs$ ILC weight matrix $\tB$ is designed to offer unit gain
to the foregrounds while minimizing the total variance of the vector estimate $\widehat{\vecf}$.
In other words, matrix $\tB$ is the minimizer of $E(\vert\vert\tB \bdy\vert\vert^2)$ under the constraint $\tB\tF=\tF$. The
weights matrix $\tB$ thus solves the following constrained variance minimization problem
\begin{equation}\label{eq:leminprob}
  \min_{\tB\tF=\tF} {\rm Tr}\left(\tB \tR \tB^t\right),  
\end{equation}
where $\tR$ is the covariance matrix of the observations $\bdy$ and ${\rm Tr}$ is the matrix trace operator.  That problem can be solved by introducing a Lagrange
multiplier matrix $\tLambda$ and the Lagrangian
\begin{equation}
  \label{qqq:lagrange} 
  \mathcal{L}(\tB, \tLambda)= 
  {\rm Tr}\left(\tB \tR \tB^t\right)
  -
  {\rm Tr}\left(\tLambda^t(\tB\tF-\tF)\right). 
\end{equation}
By differentiating (\ref{qqq:lagrange}) with respect to $\tB$, one finds that $\partial \mathcal{L}(\tB,
\tLambda)/\partial\tB = 0$ entails
\begin{equation}
  \label{qqq:semisol}
  2 \tB \tR =  \tLambda \tF^t.
\end{equation}
By imposing the constraint $\tB\tF=\tF$ on (\ref{qqq:semisol}), one then finds that $\tLambda = 2 \tF(\tF^t \tR^{-1}
\tF)^{-1}$.  Hence, the solution of the problem (\ref{eq:leminprob}) is the foreground ILC weight matrix given by
\begin{equation}
  \label{qqq:forfilter} 
  \tB = 
  \tF\left(\tF^t \tR^{-1}\tF\right)^{-1}\tF^t   \tR^{-1}.  
\end{equation}
Comparing Eq.~(\ref{qqq:forfilter}) to Eq.~(\ref{eq:ILC}), multi-dimensional ILC appears as a direct generalization of
one-dimensional ILC.\footnote{The ILC estimate of the CMB vector of emission in each frequency channel is obtained by
  applying the filter $\bda\left(\bda^t \tR^{-1}\bda\right)^{-1}\bda^t\tR^{-1}$ (i.e., the filter of eq. (\ref{eq:ILC})
  multiplied on the left by the vector $\bda$).}

One can immediately notice that expression~(\ref{qqq:forfilter}) for $\tB$ is invariant if $\tF$ is changed into $\tF \tT$
for any invertible matrix $\tT$.  Hence, as already mentioned in Section~\ref{sec:ism-subspace}, implementing the
foreground ILC filter~(\ref{qqq:forfilter}) only requires that the foreground mixing matrix $\tF$ be known up to right
multiplication by an invertible factor.  Again, in other words, the only meaningful and mandatory quantity for implementing
a multi-dimensional ILC is the column space of $\tF$.

\subsection{Estimation of the foreground subspace} \label{subsec:dim-FG}

In this section, we propose a method for estimating the foreground subspace locally, that is, in each needlet domain.  We
consider only the case where the model accounts for the CMB, an $\rfg$-dimensional foreground component and noise at a
known level:
\begin{equation}
  \label{qqq:}
  \tR = C_{\rm CMB}\bda\bda^t~+~\tF \tG \tF^t + \tR_{\bdN}  
\end{equation}
and we want to estimate the foreground subspace $\Col(\tF)$ from an
estimate $\widehat\tR$ of $\tR$.
Define the $\nobs\times(\rfg+1)$ matrix:
\begin{equation}\label{eq:defL}
  \tL = \left[ \, \bda C_{\rm CMB}^{1/2} \ \right | \left. \  \tF \tG^{1/2}   \,  \right]
\end{equation}
where the first column $\bda C_{\rm CMB}^{1/2} $ of $\tL$, containing the CMB frequency scaling vector (which is known) is distinguished from the $\rfg$ \emph{unknown} remaining columns $\tF \tG^{1/2}$ associated to the foregrounds.
So the signal part of the covariance matrix is $\tL\tL^t$:
\begin{displaymath}
  \tR = \tL\tL^t + \tR_{\bdN}.
\end{displaymath}
Our procedure for estimating the column space $\Col(\tF)$ of the foreground mixing matrix $\tF$ is in two steps.  In a first step, we obtain an estimate of $\tL$ up to right
multiplication by a rotation matrix (and an estimate for the dimension $\rfg$ of the foreground subspace) using the
knowledge of the noise covariance matrix.  In a second step, we use the fact that the first column of $\tL$ is known (up to
scale) to obtain an estimate of $\Col(\tF)$.  That is described next.

Denote the eigenvalue decomposition of the noise-whitened signal
covariance matrix $\tR_{\bdN}^{-1/2}\tL\tL^t \tR_{\bdN}^{-1/2}$ as
\begin{displaymath}
  \tR_{\bdN}^{-1/2}\tL\tL^t \tR_{\bdN}^{-1/2}
  =
  \tU\Delta\tU^t .
\end{displaymath}
where $\tU$ is orthonormal: $\tU\tU^t=\tI$, and $\Delta$ is diagonal.
Now, 
\begin{align}
  \tR_{\bdN}^{-1/2}\tR\tR_{\bdN}^{-1/2} 
  &= \tR_{\bdN}^{-1/2}(\tL\tL^t + \tR_{\bdN} ) \tR_{\bdN}^{-1/2}   \nonumber \\
  &= \tR_{\bdN}^{-1/2}\tL\tL^t \tR_{\bdN}^{-1/2} + \tI  \nonumber \\
  &= \tU\Delta \tU^t + \tU \tU^t  \nonumber \\
  &= \tU [ \Delta + \tI ] \tU^t,  \nonumber 
\end{align}
showing that $\tR_{\bdN}^{-1/2}\tR\tR_{\bdN}^{-1/2}$ and
$\tR_{\bdN}^{-1/2}\tL\tL^t\tR_{\bdN}^{-1/2}$ share the same
eigen-vectors but that the former has its eigenvalues shifted by~$1$
with respect to the latter.
Further, if $\tL$ has rank $\rfg+1$ then its eigen-structure actually is
\begin{equation}\label{eq:splitSN}
  \tU\Delta\tU^t 
  = 
  [ \tU_s \tU_n ] 
  \left[\begin{array}{cc} \Delta_{\tS}& 0\\ 0& 0\end{array}\right]
  [ \tU_s \tU_n ] ^t
\end{equation}  
where $\tU_s$ has $(\rfg+1)$ columns, $\tU_n$  has ${\nobs-(\rfg+1)}$ columns, and $\Delta_{\tS}$ is a $(\rfg+1)\times (\rfg+1)$ diagonal matrix.

\subsubsection{Estimation of $\rfg$  and $\tL$} \label{subsubsec:dim}

In the needlet domain considered, given an estimate $\widehat\tR$ of
$\tR$, we compute the eigen-decomposition:
\begin{equation}\label{eq:EVDhR}
  \tR_{\bdN}^{-1/2}\widehat{\tR}\tR_{\bdN}^{-1/2} = \widehat{\tU}\widehat{\tD}\widehat{\tU}^t  
\end{equation}
and, similar to eq.~(\ref{eq:splitSN}), we denote by
$\widehat{\tD}_{\tS}$ the sub-block of $\widehat\tD$ corresponding to
the 
eigenvalues that are larger than ${(1+\varepsilon)}$ and by
$\widehat\tU_\tS$ the corresponding subset of columns of
$\widehat\tU$. Here $\varepsilon$ is a threshold above which the
observation is not dominated by instrumental noise (see section
\ref{sec:results} for the choice of the threshold). This threshold
condition thus provides an estimate for the dimension $\rfg$ of the
foreground subspace in the needlet domain given the dimension
$(\rfg+1)$ of the sub-block $\widehat{\tD}_{\tS}$ which fulfills the
threshold condition.
 
In this preprocessing, the dimension of the estimated signal subspace,
$(\rfg+1)$, depends on the level of noise in the needlet domain
considered, so the signal subspace is estimated locally, both in space
and in scale.  This processing thus locally performs a rank reduction
of the observations covariance matrix allowing the reduction of the
instrumental noise in the reconstruction.

By construction, the matrix
\begin{equation}\label{eq:defhM}
  \widehat\tM = \tR_{\bdN}^{1/2} \widehat{\tU}_{\tS} \left(\widehat{\tD}_{\tS} -\tI\right)^{1/2}
\end{equation}
is such that $\tR_{\bdN}^{-1/2}\widehat\tM\widehat\tM^t
\tR_{\bdN}^{-1/2}$ is close to $\tR_{\bdN}^{-1/2}\tL\tL^t
\tR_{\bdN}^{-1/2}$ if $\widehat\tR$ is close to $\tR$.  
That property, in turn, implies that $\widehat\tM \tO$ is close to
$\tL$ for some (undetermined) rotation matrix $\tO$, completing the
first step of our estimation procedure.

\subsubsection{Estimation of the foreground column space $\Col(\tF)$}\label{subsubsec:col-F}

In the second step, we note that the rotation matrix $\tO$ should be
such that $\widehat\tM \tO$ is close to $\tL$.  However, only the
first column of $\tL$ is known, up to scale.  Hence, we partition
$\tO$ as $\tO = [\, \bdv\, |\, \tV]$ where $\bdv$ is a unit norm
vector and $\tV$ is an $(\rfg+1)\times\rfg$ matrix.  The only
available constraint is thus that $\widehat\tM\bdv$ should be close to
the first column of $\tL$.  However, we cannot expect to find a $\bdv$
such that $\widehat\tM\bdv$ is strictly equal to $\bda C_{\rm
  CMB}^{1/2}$ because $\widehat\tM$ is estimated from the data so that
$\Col(\widehat\tM)$ does not necessarily contain $\bda$ (as would be
the case for $\widehat\tR=\tR$).
The best we can do is to determine $\bdv$ such that $\widehat\tM\bdv$
is equal to the \emph{projection} of $\bda C_{\rm CMB}^{1/2}$ onto
$\Col(\widehat\tM)$.  The orthogonal projection matrix is
$\widehat\tM(\widehat\tM^t \widehat\tM)\inv \widehat\tM^t$ so that the
projection is
\begin{displaymath}
  \widehat\tM \left(\widehat\tM^t \widehat\tM\right)\inv \widehat\tM^t\ \bda C_{\rm CMB}^{1/2}  .
\end{displaymath}
Let us then denote by $\widetilde{\bda}$ the vector
\begin{equation}\label{eq:defbvec}
  \widetilde{\bda} =  \left(\widehat\tM^t \widehat\tM\right)\inv \widehat\tM^t\ \bda .
\end{equation}
The projection of $\bda C_{\rm CMB}^{1/2}$ onto $\Col(\widehat\tM)$
then is $\widehat\tM\, \widetilde{\bda} \, C_{\rm CMB}^{1/2} $ and vector $\bdv$
is therefore given by $\bdv = \widetilde{\bda}\, C_{\rm CMB}^{1/2}$.  Recall that
$\bdv$ is a unit norm vector so we must have:
\begin{equation}\label{eq:findv}
  \bdv = \widetilde{\bda} / |\widetilde{\bda}|
  \qquad\text{and}\quad
  C_{\rm CMB} = 1/|\widetilde{\bda}|^2 .
\end{equation}
Once vector $\bdv$ is determined, the constraint that $\tO = [\,
\bdv\, |\, \tV]$ is a rotation matrix uniquely determines $\tV$ up to
right multiplication by a rotation factor.  However, nothing more is
required to determine the foreground subspace, as already stressed.
Our procedure is therefore complete and can be summarized by the
following steps:
\begin{itemize}
\item Compute the eigen-decomposition (\ref{eq:EVDhR}) of the noise
  whitened covariance matrix. Obtain an estimate of $\rfg$ from comparing the level of the eigenvalues to the noise level.
\item Form matrix $\widehat\tM$ by eq.~(\ref{eq:defhM}), compute
  vector $\widetilde{\bda}$ by eq.~(\ref{eq:defbvec}) and get $\bdv$ by
  normalization~(\ref{eq:findv}).
\item Compute an $(\rfg+1)\times \rfg$ matrix $\tV$ such that matrix
  $[\bdv \,|\, \tV]$ is a rotation.
\item Obtain a basis of the foreground subspace as $\widehat\tF =
  \widehat \tM\, \tV$.
\item Compute the $\rfg$-dimensional ILC filter
  \begin{displaymath}
    \widehat\tB 
    = \widehat{\tF}\left(\widehat{\tF}^t \tR^{-1}\widehat{\tF}\right)^{-1}\widehat{\tF}^t \tR^{-1}
  \end{displaymath}
\end{itemize}

\begin{figure*}
  \begin{center}
    \includegraphics[width=5.3cm]{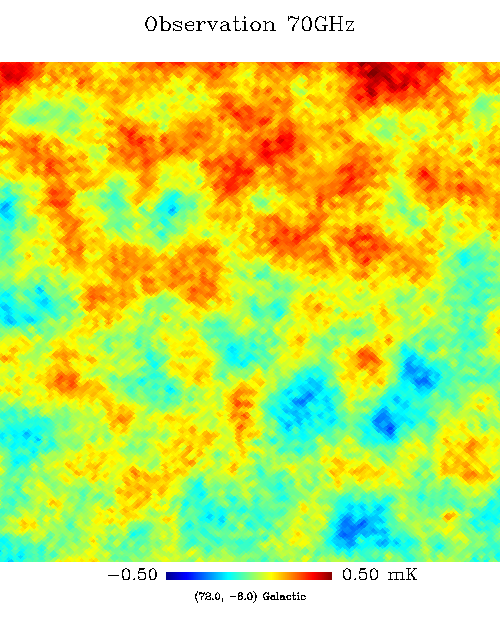}
    \includegraphics[width=5.3cm]{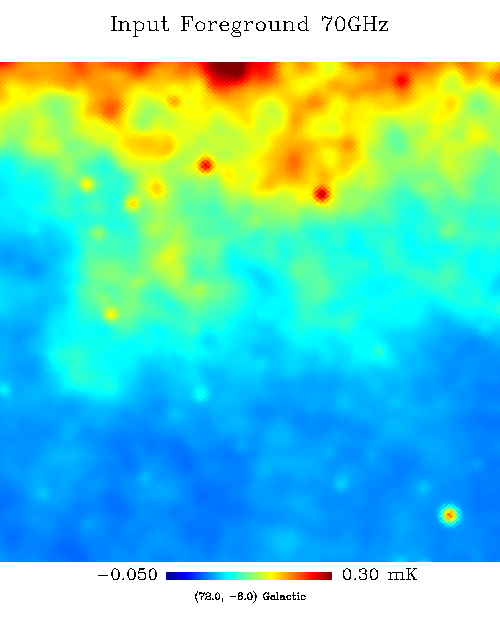}
    \includegraphics[width=5.3cm]{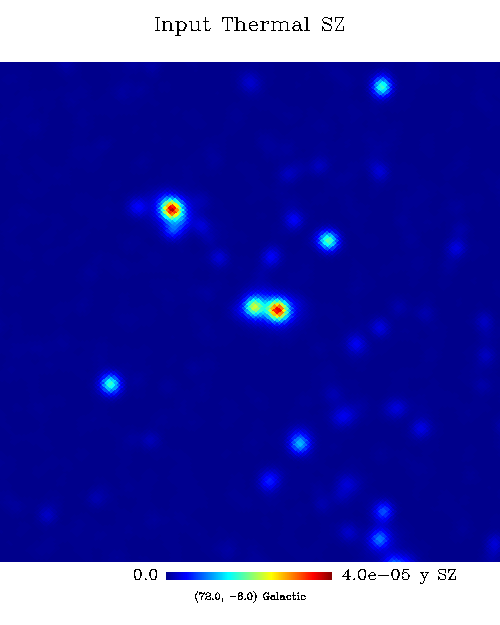}
  \end{center}
\caption{{\bf Simulated Planck observations.} A $12.5^\circ\times 12.5^\circ$ patch of the simulated sky located at low Galactic latitude, around Galactic coordinates of $(l,b) = (72^\circ,-8^\circ)$. From left to right: observed map, foreground map, and thermal SZ map  at 70 GHz. All maps are at the resolution of the 70 GHz channel (14 arc-minutes).}
\label{Fig:frgdIn}
\end{figure*}

\subsection{Projecting foregrounds orthogonally to both thermal SZ and CMB }\label{subsubsec:sz-orth}

The foreground multidimensional ILC filter can be generalised further.
Thermal SZ emission can be singled out in the same way as the CMB, in
which case we may require that there is no thermal SZ residual in the
reconstructed foreground map. This is doable because the emission law
of the SZ component, like that of the CMB, is known.  We then write
the model of emissions as:
\begin{equation}
  \label{qqq:modelsz}
  \tA = \left[
\begin{array}{ccc}
\bda & \bdb & \tF
\end{array}\right],\qquad 
\bdx=\left[
\begin{array}{c}      
s \\
z \\
\bdg 
\end{array}\right].
\end{equation}
where we have explicitly distinguished the thermal SZ emission $z$
from the other foregrounds through its frequency scaling vector
$\vecb$ (emission law). We may then generalize the processing
developed in Sec.~\ref{subsec:dim-FG}.
In that spirit, the FG mixing matrix $\tF$ can then be estimated in
the needlet domain considered from the set of $\rfg$ columns
orthogonal to both the projection of the CMB scaling vector and the
projection of the thermal SZ scaling vector onto the estimated
$(\rfg+2)$--dimensional signal subspace.  This guarantees that the
foreground map reconstructed by the multidimensional ILC now contains
neither SZ nor CMB (with, however, the usual caveat that the
statistics used to compute the covariance matrices must be accurate
enough).  In addition, the rank-reduction procedure (restriction of
the observations to the ${(\rfg+2)}$-dimensional signal subspace) in
each needlet region reduces the level of instrumental noise locally in
the reconstructed foregrounds.

\subsection{Discussion of special cases}

\subsubsection{Less channels than foreground dimension}

In the discussion above, we assumed that the signal subspace is the
direct sum of two subspaces: the CMB subspace which is one-dimensional
(because of the rigid scaling of the CMB with frequency) and the
foreground subspace which is $\rfg$-dimensional.
The former is not included in the latter if no combination of
foreground emission has the same scaling as the CMB across available
frequencies.  Of course, this property requires enough properly chosen
frequency channels.

When there are more components than observations, then unless the
foreground emissions are either fully correlated or very faint (below
noise), then we have ${\rfg+1>\nobs}$ and the CMB cannot be perfectly
separated from the foregrounds.
When there are enough independent observations (\textit{i.e.} a large
number of channels), then ${\rfg<\nobs}$, and in general the CMB
subspace is not contained in the (larger) foreground subspace.
Separation is then possible up to finite-sample size errors in the
determination of the appropriate subspaces.

Note in the passing that the kinetic SZ cannot be separated from the
CMB on the basis of its emission law.  Throughout this paper the CMB,
distinguished solely by its known emission law $\bda$, also includes
the kinetic SZ effect.

\begin{figure}
   \begin{center}
     \includegraphics[width=7.5cm]{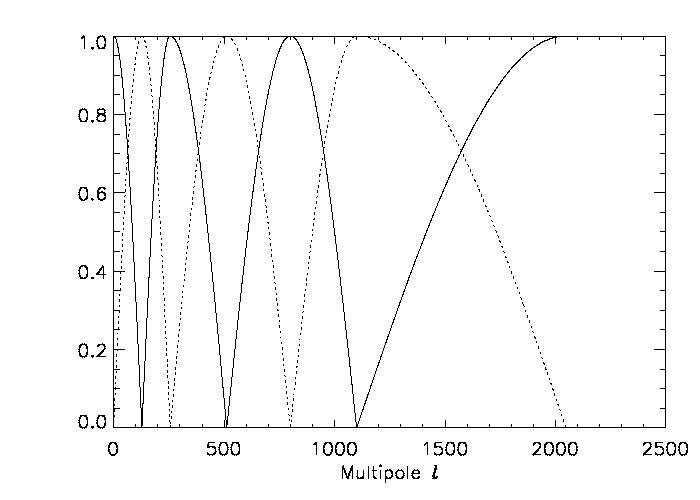}
   \end{center}
 \caption{The spectral bands used in this work for the definition of the needlets.}
 \label{Fig:bands}
 \end{figure}

\begin{figure*}
  \begin{center}
    \includegraphics[width=5.3cm]{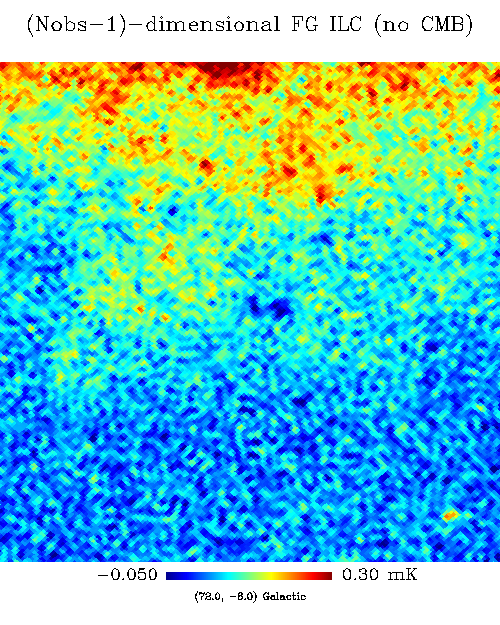}
    \includegraphics[width=5.3cm]{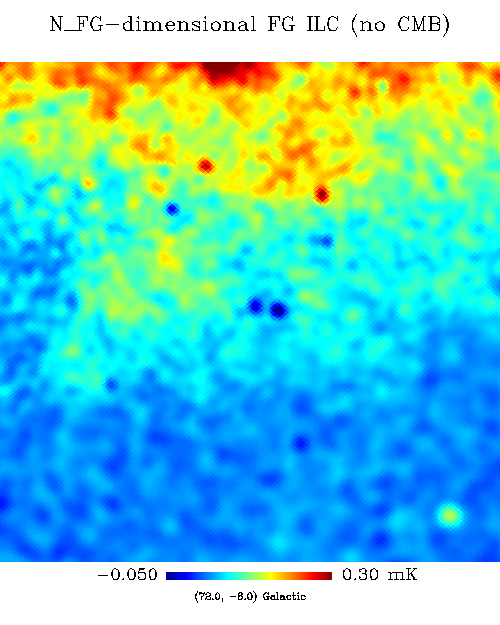}
    \includegraphics[width=5.3cm]{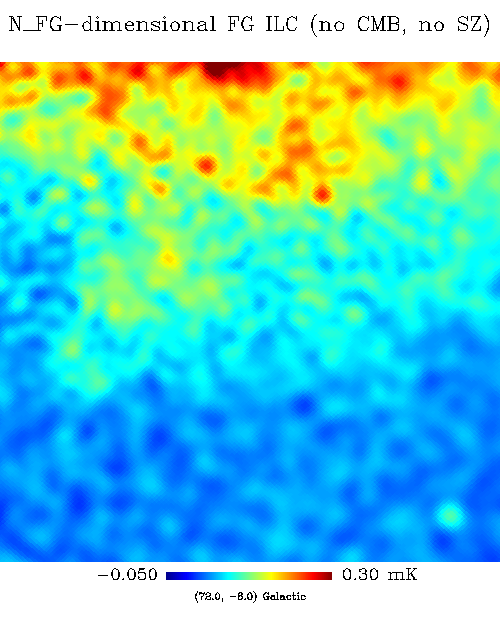}
    \includegraphics[width=5.3cm]{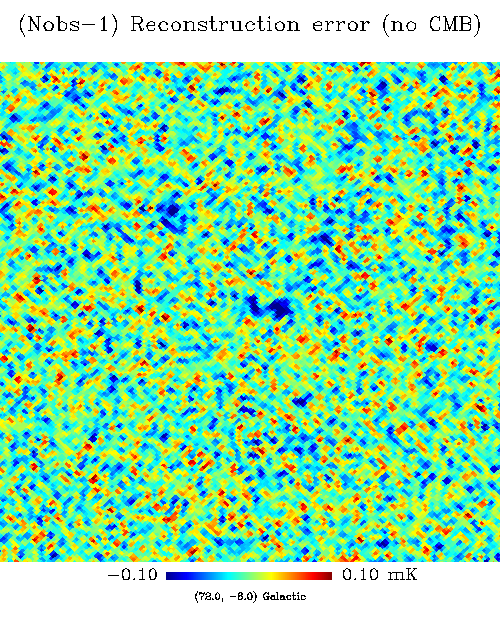}
    \includegraphics[width=5.3cm]{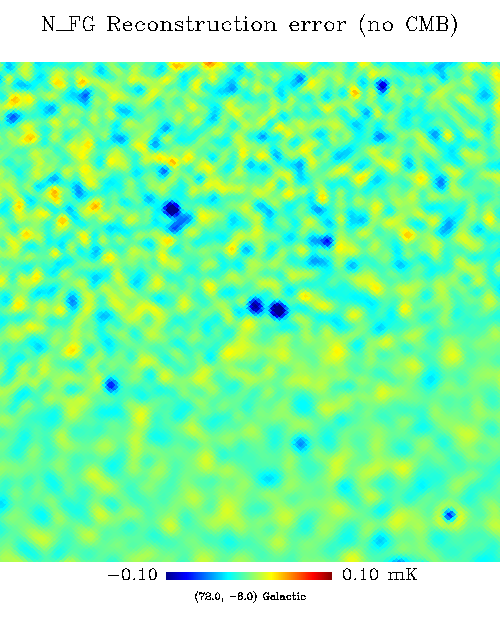}
    \includegraphics[width=5.3cm]{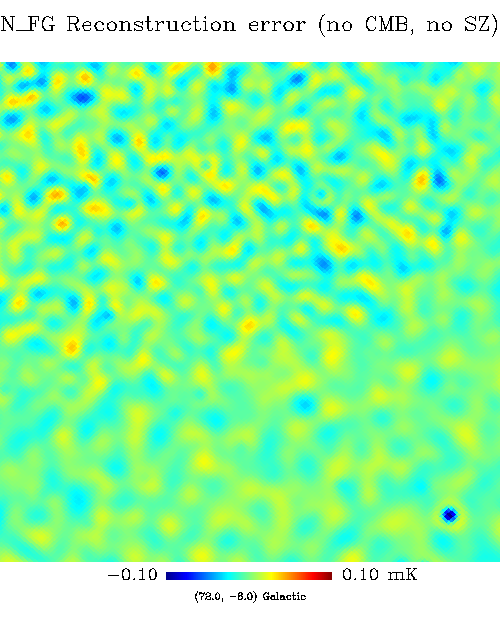}
  \end{center}
\caption{{\bf 70GHz foreground multidimensional ILC reconstruction at low Galactic latitude around $\boldsymbol{(l,b) = (72^\circ,-8^\circ)}$}. Top: CMB-orthogonal $(\nobs-1)$--dimensional ILC map, CMB-orthogonal $\rfg$--dimensional ILC map, (CMB+SZ)-orthogonal $\rfg$--dimensional ILC map. Bottom: error (difference input-output) maps respectively for CMB-orthogonal $(\nobs-1)$--dimensional ILC, CMB-orthogonal $\rfg$--dimensional ILC, and (CMB+SZ)-orthogonal $\rfg$--dimensional ILC.}
\label{Fig:frgdILC}
\end{figure*}

\begin{figure*}
  \begin{center}
    \includegraphics[width=8.5cm]{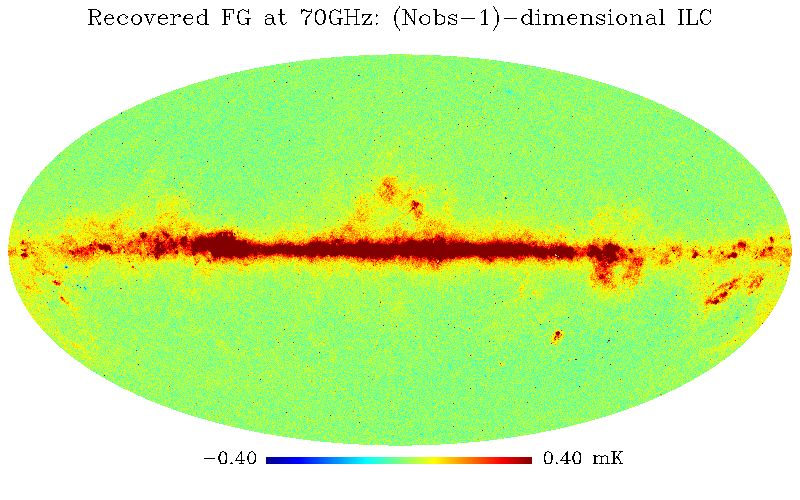}
    \includegraphics[width=8.5cm]{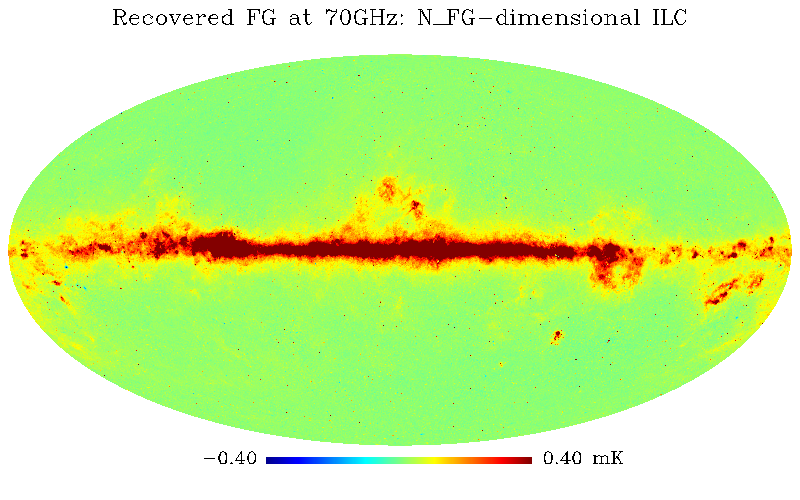}
    \includegraphics[width=8.5cm]{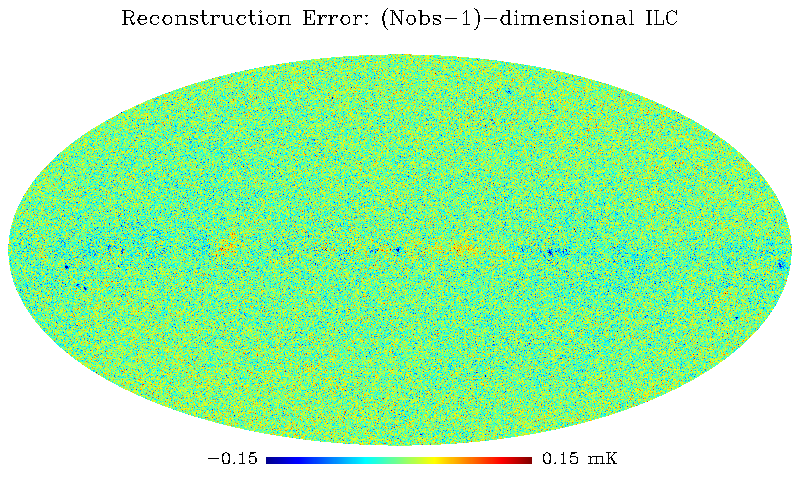}
    \includegraphics[width=8.5cm]{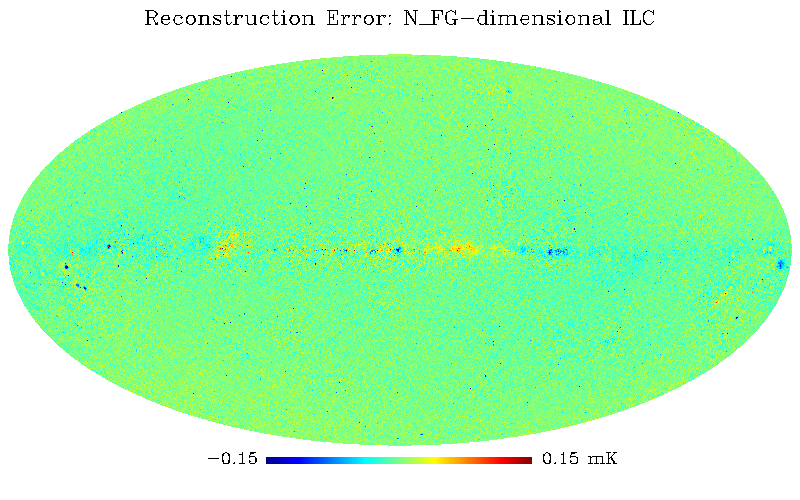}
  \end{center}
  \caption{{\bf Foregrounds reconstruction at 70 GHz.} Left panels: $(\nobs-1)$--dimensional needlet ILC output map and
      needlet error (difference input-output) map. Right panels: $\rfg$--dimensional needlet ILC output map and
      needlet error (difference
      input-output) map. The $\rfg$--dimensional needlet ILC guarantees the reduction of the noise contamination.}
  \label{Fig:outputs}
\end{figure*}

\subsubsection{CMB subtraction as an $(\nobs-1)$--dimensional
  ILC}\label{subsubsec:n-1}

A foreground estimation has been obtained in \cite{ghosh10} by
subtracting the CMB-ILC estimate from the observations data,
${\widehat{\vecf} = \bdy - \veca \widehat{s}}$. It is interesting to
note that the CMB subtraction procedure is equivalent to an
${(\nobs-1)}$--dimensional ILC filtering, \textit{i.e.} the particular
multidimensional ILC filtering where the dimension of the foreground
subspace is assumed to be constant over the whole sky and the whole
range of scales, and equal to ${(\nobs-1)}$.  Indeed, the CMB
subtracted estimate expands as follows
\begin{align}
  \widehat{\vecf} 
  &= \bdy - \veca \frac{\veca^t \, \tR^{-1}\bdy}{\veca^t \, \tR^{-1} \, \veca}, \nonumber \\
  &=  \tW^{-1}\left(\mI - \tW\veca\left(\left(\tW\veca\right)^t
      \tW\veca\right)^{-1}\left(\tW\veca\right)^t\right)\tW\bdy,\nonumber\\ 
  &= \tW^{-1}\left(\mI - \tP_1\right)\tW\bdy,\label{qqq:subtraction}
\end{align} 
where $\tW = \tR^{-1/2}$ denotes the inverse square root of the data covariance matrix.

Matrix ${\tP_1= \tW\veca\left(\left(\tW\veca\right)^t \tW\veca\right)^{-1}\left(\tW\veca\right)^t}$ is an \emph{orthogonal}
projection (${\tP_1^2 = \tP_1}$ and ${\tP_1^t = \tP_1}$) onto the line ${\Span\left(\tW\bda\right)}$ (one-dimensional
`whitened' CMB subspace). It implies that ${\mI - \tP_1 = \tP_\mathcal{H}}$ is the projection onto the
${(\nobs-1)}$--dimensional hyperplane ${\mathcal{H} = \left[\Span\left(\tW\bda\right)\right]^\perp}$, which is orthogonal
and complementary to the one-dimensional whitened CMB subspace. Let us denote ${\bdv = \tW\bda / |\tW\bda|}$ and consider
an ${\nobs\times (\nobs-1)}$ matrix $\tV$ such that ${[\bdv \,|\, \tV]}$ is a rotation. Then ${\tP_\mathcal{H} = \tV
  \left(\tV^t \tV\right)^{-1}\tV^t}$ and, by denoting $\tF = \tW^{-1}\tV$, we get
\begin{align}
  \widehat{\vecf} 
  &= \tW^{-1}\tP_\mathcal{H}\tW\bdy,\nonumber\\
  &=  \tW^{-1}\tV \left(\tV^t \tV\right)^{-1}\tV^t\tW\bdy,\nonumber\\
  &=   \tF \left(\tF^t\tR^{-1}\tF\right)^{-1}\tF^t\tR^{-1}\bdy,  \label{qqq:subtraction2}
\end{align} 
which completes the proof since $\tF$ is full rank ${(\nobs-1)}$. 
Here, it is interesting to notice that the ${(\nobs-1)}$--dimensional ILC can be obtained \emph{without even knowing} the
mixing matrix $\tF$ since the procedure becomes equivalent to the estimation obtained by subtracting the CMB-ILC estimate
from the observations data.

This equivalence means that the CMB subtraction procedure does not take advantage of the fact that the foreground mixing
matrix can be almost rank-deficient in some regions of the sky or at some scales (for instance at small scale where the
instrumental noise is dominant). The ${(\nobs-1)}$--dimensional subspace $\Col(\tF)$ reconstructed here thus includes both
noise and foregrounds components. Consequently, such a foreground reconstruction is noisy.
The $\rfg$--dimensional ILC procedure described in section~\ref{subsec:dim-FG} performs a cleaner foreground reconstruction
(in terms of signal to noise ratio) because the effective rank $\rfg$ of the foreground subspace and the foreground mixing
matrix (with reduced rank) are estimated locally in each needlet domain. In effect, this boils down to performing at the
same time both component separation, and denoising by thresholding the needlet coefficients.

\begin{figure}
   \begin{center}
     \includegraphics[width=7.5cm]{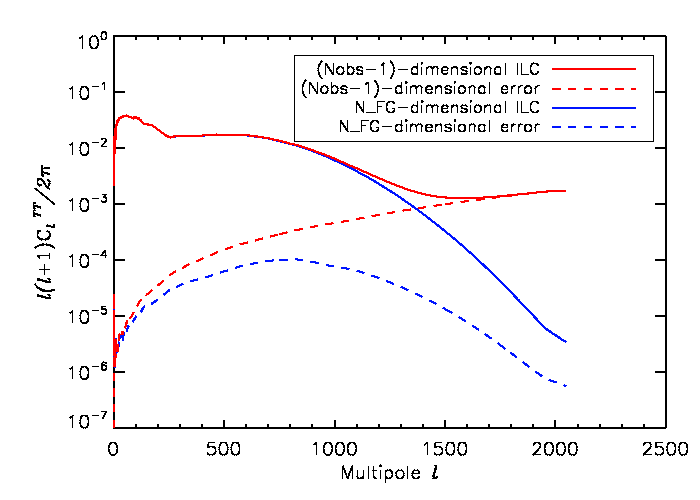}
   \end{center}
 \caption{{\bf Power spectrum of the recovered foregrounds at 70 GHz and of the ILC reconstruction error:} {$(\nobs-1)$--dimensional} foreground ILC (solid red) and error (dashed red), $\rfg$--dimensional foreground ILC (solid blue) and error (dashed blue). We clearly see the suppression of the noise at high $\ell$ for the $\rfg$--dimensional foreground ILC reconstruction.}
 \label{Fig:errspec2}
 \end{figure}

\section{Planck simulations and results}\label{sec:results}

\begin{figure*}
   \begin{center}
     \includegraphics[width=10cm]{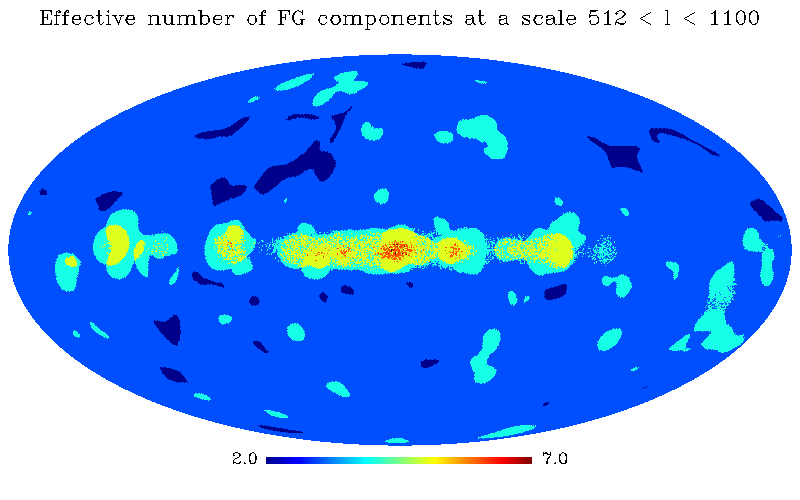}
     \includegraphics[width=5cm]{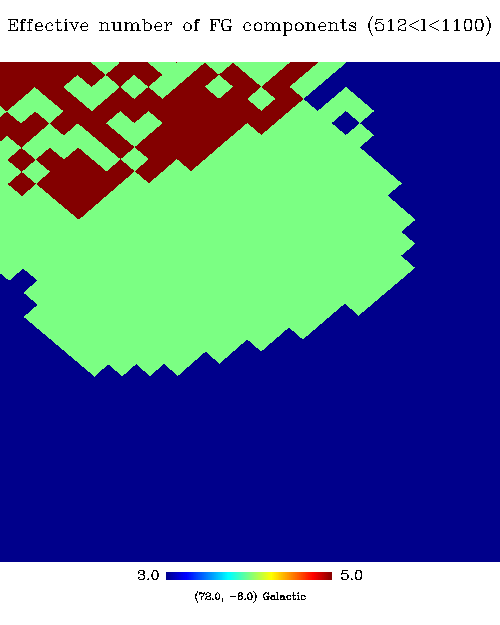}
   \end{center}
 \caption{Left: Effective number $\rfg$ of foreground components (effective rank of the foreground covariance matrix) at a scale $512 < \ell < 1100$ estimated in each needlet domain from the $80\%$ noise threshold. Components contributing less than $20\%$ of the total observation are thus neglected in this analysis (Note that at this scale the original number of useful channels is eight instead of nine because the $30 GHz$ channel does not have enough resolution). Right: same at low Galactic latitude around $(l,b) = (72^\circ,-8^\circ)$.}
 \label{Fig:nfg}
 \end{figure*}

We now turn to illustrating this discussion with examples based on simulated data sets.  We apply our multidimensional FG
ILC filter on a frame of needlets. The spectral bands used in the definition of the needlets are shown in figure \ref{Fig:bands}. For each needlet domain considered, we both project the data onto the `full rank'
foreground subspace (equivalent to simple CMB-ILC subtraction), and onto a `reduced rank' foreground subspace. For the
latter, we reject the eigenvalues of the covariance matrix of the observation needlet coefficients smaller than $1.25$ times the noise covariance level, i.e. values for which the
instrumental noise contributes more than $80\%$ of the total emission. This is a somewhat arbitrary, but reasonable choice, chosen here for illustration. In practice, this threshold can be fixed more rigorously, considering the trade-off between rejecting low-level
foreground emission, and letting in the final map too much instrumental noise.

As a refinement the number of relevant foreground components could be estimated without even imposing any arbitrary threshold, e.g. by using the Akaike Information Criterium \citep{1974ITAC...19..716A}. This criterium consists in maximizing the likelihood of the observations given the model, taking into account a particular penalty imposed on the number of free parameters entering in the model (e.g. the number of foreground components, or equivalently the rank of the foreground mixing matrix).  


Our investigations are carried out on sky temperature simulations generated with the Planck Sky Model (PSM) version 1.6.6. Sky
simulations include Gaussian CMB generated assuming a theoretical angular spectrum fitting the WMAP observations, thermal
and kinetic SZ effect, four components of Galactic ISM emission including thermal and spinning dust, synchrotron, and
free-free, and emission from point sources (radio and infrared). The resolution and noise level of the observations
correspond to nominal mission parameters as described in the Planck ``Blue Book''. Nine frequency channels are used 
in this simulation and correspond to the Planck HFI and LFI channels. Details about PSM simulations can be
found in \cite{2008A&A...491..597L} and \cite{2009A&A...503..691B}.

Figure~\ref{Fig:frgdIn} shows the `observed' 70 GHz map, the input foreground map at 70 GHz, and the thermal SZ map, all at
the resolution of the 70~GHz channel.  Our maps are centred on an interesting region which is both at low Galactic
latitude, around $(l,b) = (72^\circ,-8^\circ)$ and close to a set of bright galaxy clusters.  The 70 GHz reconstructed
foregrounds, recovered by multidimensional ILC filtering, are shown in the same region of the sky on the top panels of
figure~\ref{Fig:frgdILC}.  The corresponding reconstruction error (difference between reconstructed output and original
input) is displayed on the bottom row of the same figure. The $(\nobs-1)$--dimensional ILC reconstruction (left panels),
equivalent to a simple subtraction of the ILC estimate of the CMB map to the observation map, is clearly noisy (as expected
from the discussion of section~\ref{subsubsec:n-1}). The reduction of the noise in the foreground reconstruction is
achieved by performing a $\rfg$--dimensional ILC reconstruction (middle panels), where $\rfg$ is the local dimension of the
FG subspace depending both on the needlet scale and on the pixel.
We observe the leakage of a thermal SZ emission in the FG reconstruction on the left and middle panels of
figure~\ref{Fig:frgdILC}.  Using the modified `reduced rank' ILC introduced in section~\ref{subsubsec:sz-orth}, we obtain
instead the reconstruction displayed on the right panels of figure~\ref{Fig:frgdILC} with no visible contamination by SZ
emission.

For completeness the same results are shown on full sky maps in figure~\ref{Fig:outputs} and we have plotted the
corresponding power spectra in figure~\ref{Fig:errspec2}. The suppression of the noise contamination is clearly visible on
the spectrum at high $\ell$ when a $\rfg$--dimensional ILC method is employed.

Figure \ref{Fig:nfg} shows, for the fifth needlet band (scales comprised between $\ell = 512$ and $\ell = 1100$), the effective number $\rfg$ of foreground components (i.e. the effective rank of the foreground covariance matrix) which has been estimated in each needlet domain from the $80\%$ noise threshold (foreground components contributing less than $20\%$ of the total emission in the needlet domain have been neglected). 
On the right panel of the figure, a $12.5^\circ\times 12.5^\circ$ patch of sky at low Galactic latitude, centred around $(l,b) = (72^\circ,-8^\circ)$, explicitely shows the effective number of foreground components estimated in this region. This number decreases according to the distance to the Galactic plane. This is consistent with the bottom right panel of figure \ref{Fig:frgdILC} showing that the residual noise is locally distributed and decreases according to the distance to the Galactic plane.

\section{Conclusion}

In this article, we have shown how the standard ILC procedure, originally dedicated to the CMB extraction, can be extended
for the reconstruction of complex astrophysical emissions, beyond the CMB alone.  We have developed generalised ILC filters
(\emph{multidimensional ILC}) to reconstruct the diffuse emission of a complex multidimensional component originating from
multiple correlated emissions, such as the total Galactic foreground emission.  
Similar, though pixel-based extensions have been also implemented in a
fastICA-based code, AltICA,  as used in \citet{2008A&A...491..597L} and are integrated in the
Planck LFI Data Processing Center pipeline (C. Baccigalupi, R.Stompor,
private communication).
Our estimators were implemented on a
needlet frame and tested on simulations of Planck observations. This new ILC filtering successfully reconstructs the
foreground emission, exempt from both the CMB and the SZ emission, and with a reduced level of instrumental noise.

\section*{Acknowledgements}
\thanks{ We thank Tuhin Ghosh, Radek Stompor and
  Carlo Baccigalupi for useful conversations related to this work.  Some of the results in this paper have been
  derived using the HEALPix package \citep{gorski05}.  We also acknowledge the use of the Planck Sky Model, developed by
  the Component Separation Working Group (WG2) of the Planck Collaboration.}

\bibliography{generalised_ilc}

\begin{thebibliography}{30}
\expandafter\ifx\csname natexlab\endcsname\relax\def\natexlab#1{#1}\fi

\bibitem[{{Akaike}(1974)}]{1974ITAC...19..716A}
{Akaike} H., 1974, IEEE Transactions on Automatic Control, 19, 716

\bibitem[{{Bennett} {et~al}\mbox{.}(2003){Bennett}, {Hill}, {Hinshaw}, {Nolta},
  {Odegard}, {Page}, {Spergel}, {Weiland}, {Wright}, {Halpern}, {Jarosik},
  {Kogut}, {Limon}, {Meyer}, {Tucker}, \& {Wollack}}]{2003ApJS..148...97B}
{Bennett} C.~L. {et~al.}, 2003, \apjs, 148, 97

\bibitem[{{Betoule} {et~al}\mbox{.}(2009){Betoule}, {Pierpaoli},
  {Delabrouille}, {Le Jeune}, \& {Cardoso}}]{2009A&A...503..691B}
{Betoule} M., {Pierpaoli} E., {Delabrouille} J., {Le Jeune} M., {Cardoso} J.,
  2009, \aap, 503, 691

\bibitem[{{Bobin} {et~al}\mbox{.}(2008){Bobin}, {Moudden}, {Starck}, {Fadili},
  \& {Aghanim}}]{2008StMet...5..307B}
{Bobin} J., {Moudden} Y., {Starck} J., {Fadili} J., {Aghanim} N., 2008,
  Statistical Methodology, 5, 307

\bibitem[{{Bonaldi} {et~al}\mbox{.}(2006){Bonaldi}, {Bedini}, {Salerno},
  {Baccigalupi}, \& {de Zotti}}]{2006MNRAS.373..271B}
{Bonaldi} A., {Bedini} L., {Salerno} E., {Baccigalupi} C., {de Zotti} G., 2006,
  \mnras, 373, 271

\bibitem[{{Bouchet} \& {Gispert}(1999)}]{1999NewA....4..443B}
{Bouchet} F.~R., {Gispert} R., 1999, New Astronomy, 4, 443

\bibitem[{{Cardoso}(1998)}]{JADE}
{Cardoso} J.-F., 1998, Proceedings of the IEEE, 9, 2009

\bibitem[{{Cardoso} {et~al}\mbox{.}(2008){Cardoso}, {Le Jeune}, {Delabrouille},
  {Betoule}, \& {Patanchon}}]{2008ISTSP...2..735C}
{Cardoso} J.-F., {Le Jeune} M., {Delabrouille} J., {Betoule} M., {Patanchon}
  G., 2008, IEEE Journal of Selected Topics in Signal Processing, vol.~2, issue
  5, pp.~735-746, 2, 735

\bibitem[{{Delabrouille} \& {Cardoso}(2009)}]{2009LNP...665..159D}
{Delabrouille} J., {Cardoso} J., 2009, in Lecture Notes in Physics, Vol. 665,
  Berlin Springer Verlag, {V.~J.~Martinez, E.~Saar, E.~M.~Gonzales, \&
  M.~J.~Pons-Borderia}, ed., pp. 159--+

\bibitem[{{Delabrouille} {et~al}\mbox{.}(2009){Delabrouille}, {Cardoso}, {Le
  Jeune}, {Betoule}, {Fay}, \& {Guilloux}}]{2009A&A...493..835D}
{Delabrouille} J., {Cardoso} J., {Le Jeune} M., {Betoule} M., {Fay} G.,
  {Guilloux} F., 2009, \aap, 493, 835

\bibitem[{{Delabrouille}, {Cardoso} \& {Patanchon}(2003){Delabrouille},
  {Cardoso}, \& {Patanchon}}]{2003MNRAS.346.1089D}
{Delabrouille} J., {Cardoso} J., {Patanchon} G., 2003, \mnras, 346, 1089

\bibitem[{{Delabrouille}, {Patanchon} \& {Audit}(2002){Delabrouille},
  {Patanchon}, \& {Audit}}]{2002MNRAS.330..807D}
{Delabrouille} J., {Patanchon} G., {Audit} E., 2002, \mnras, 330, 807

\bibitem[{{Dick}, {Remazeilles} \& {Delabrouille}(2010){Dick}, {Remazeilles},
  \& {Delabrouille}}]{2010MNRAS.401.1602D}
{Dick} J., {Remazeilles} M., {Delabrouille} J., 2010, \mnras, 401, 1602

\bibitem[{{Eriksen} {et~al}\mbox{.}(2004){Eriksen}, {Banday}, {G{\'o}rski}, \&
  {Lilje}}]{2004ApJ...612..633E}
{Eriksen} H.~K., {Banday} A.~J., {G{\'o}rski} K.~M., {Lilje} P.~B., 2004, \apj,
  612, 633

\bibitem[{Fa\"y {et~al}\mbox{.}(2008)Fa\"y, Guilloux, Betoule, Cardoso,
  Delabrouille, \& Le~Jeune}]{PhysRevD.78.083013}
Fa\"y G., Guilloux F., Betoule M., Cardoso J.-F., Delabrouille J., Le~Jeune M.,
  2008, Phys. Rev. D, 78, 083013

\bibitem[{{Ghosh} {et~al}\mbox{.}(2010){Ghosh}, {Delabrouille}, {Remazeilles},
  {Cardoso}, \& {Souradeep}}]{ghosh10}
{Ghosh} T., {Delabrouille} J., {Remazeilles} M., {Cardoso} J., {Souradeep} T.,
  2010, submitted to \mnras

\bibitem[{{G{\'o}rski} {et~al}\mbox{.}(2005){G{\'o}rski}, {Hivon}, {Banday},
  {Wandelt}, {Hansen}, {Reinecke}, \& {Bartelmann}}]{gorski05}
{G{\'o}rski} K.~M., {Hivon} E., {Banday} A.~J., {Wandelt} B.~D., {Hansen}
  F.~K., {Reinecke} M., {Bartelmann} M., 2005, \apj, 622, 759

\bibitem[{Guilloux, Fa\"y \& Cardoso(2009)Guilloux, Fa\"y, \&
  Cardoso}]{guilloux:fay:cardoso:2008}
Guilloux F., Fa\"y G., Cardoso J.-F., 2009, Appl. Comput. Harmon. Anal., 26,
  143

\bibitem[{{Hobson} {et~al}\mbox{.}(1998){Hobson}, {Jones}, {Lasenby}, \&
  {Bouchet}}]{1998MNRAS.300....1H}
{Hobson} M.~P., {Jones} A.~W., {Lasenby} A.~N., {Bouchet} F.~R., 1998, \mnras,
  300, 1

\bibitem[{{Hyvarinen}(1999)}]{FastICA}
{Hyvarinen} A., 1999, IEEE Transactions on Neural Networks, 10, 626

\bibitem[{{Kim}, {Naselsky} \& {Christensen}(2008){Kim}, {Naselsky}, \&
  {Christensen}}]{2008arXiv0803.1394K}
{Kim} J., {Naselsky} P., {Christensen} P.~R., 2008, ArXiv e-prints, 803

\bibitem[{{Leach} {et~al}\mbox{.}(2008){Leach}, {Cardoso}, {Baccigalupi},
  {Barreiro}, {Betoule}, {Bobin}, {Bonaldi}, {Delabrouille}, {de Zotti},
  {Dickinson}, {Eriksen}, {Gonz{\'a}lez-Nuevo}, {Hansen}, {Herranz}, {Le
  Jeune}, {L{\'o}pez-Caniego}, {Mart{\'{\i}}nez-Gonz{\'a}lez}, {Massardi},
  {Melin}, {Miville-Desch{\^e}nes}, {Patanchon}, {Prunet}, {Ricciardi},
  {Salerno}, {Sanz}, {Starck}, {Stivoli}, {Stolyarov}, {Stompor}, \&
  {Vielva}}]{2008A&A...491..597L}
{Leach} S.~M. {et~al.}, 2008, \aap, 491, 597

\bibitem[{{Maino} {et~al}\mbox{.}(2002){Maino}, {Farusi}, {Baccigalupi},
  {Perrotta}, {Banday}, {Bedini}, {Burigana}, {De Zotti}, {G{\'o}rski}, \&
  {Salerno}}]{2002MNRAS.334...53M}
{Maino} D. {et~al.}, 2002, \mnras, 334, 53

\bibitem[{{Marinucci} {et~al}\mbox{.}(2008){Marinucci}, {Pietrobon}, {Balbi},
  {Baldi}, {Cabella}, {Kerkyacharian}, {Natoli}, {Picard}, \&
  {Vittorio}}]{2008MNRAS.383..539M}
{Marinucci} D. {et~al.}, 2008, \mnras, 383, 539

\bibitem[{{Park}, {Park} \& {Gott}(2007){Park}, {Park}, \&
  {Gott}}]{2007ApJ...660..959P}
{Park} C.-G., {Park} C., {Gott} J.~R.~I., 2007, \apj, 660, 959

\bibitem[{{Pietrobon}, {Balbi} \& {Marinucci}(2006){Pietrobon}, {Balbi}, \&
  {Marinucci}}]{2006PhRvD..74d3524P}
{Pietrobon} D., {Balbi} A., {Marinucci} D., 2006, \prd, 74, 043524

\bibitem[{{Planck Collaboration} {et~al}\mbox{.}(2011){Planck Collaboration},
  {Ade}, {Aghanim}, {Arnaud}, {Ashdown}, {Aumont}, {Baccigalupi}, {Balbi},
  {Banday}, {Barreiro}, {Bartlett}, {Battaner}, {Benabed}, {Benoit}, {Bernard},
  {Bersanelli}, {Bhatia}, {Blagrave}, {Bock}, {Bonaldi}, {Bonavera}, {Bond},
  {Borrill}, {Bouchet}, {Bucher}, {Burigana}, {Cabella}, {Cardoso}, {Catalano},
  {Cayon}, {Challinor}, {Chamballu}, {Chiang}, {Chiang}, {Christensen},
  {Clements}, {Colombi}, {Couchot}, {Coulais}, {Crill}, {Cuttaia}, {Danese},
  {Davies}, {Davis}, {de Bernardis}, {de Gasperis}, {de Rosa}, {de Zotti},
  {Delabrouille}, {Delouis}, {Desert}, {Dole}, {Donzelli}, {Dore}, {Dorl},
  {Douspis}, {Dupac}, {Efstathiou}, {Ensslin}, {Eriksen}, {Finelli}, {Forni},
  {Fosalba}, {Frailis}, {Franceschi}, {Galeotta}, {Ganga}, {Giard}, {Giardino},
  {Giraud-Heraud}, {Gonzalez-Nuevo}, {Gorski}, {Grain}, {Gratton}, {Gregorio},
  {Gruppuso}, {Hansen}, {Harrison}, {Helou}, {Henrot-Versille}, {Herranz},
  {Hildebrandt}, {Hivon}, {Hobson}, {Holmes}, {Hovest}, {Hoyland},
  {Huffenberger}, {Jaffe}, {Jones}, {Juvela}, {Keihanen}, {Keskitalo},
  {Kisner}, {Kneissl}, {Knox}, {Kurki-Suonio}, {Lagache}, {Lamarre}, {Lasenby},
  {Laureijs}, {Lawrence}, {Leach}, {Leonardi}, {Leroy}, {Lilje},
  {Linden-Vornle}, {Lockman}, {Lopez-Caniego}, {Lubin}, {Macias-Perez},
  {MacTavish}, {Maffei}, {Maino}, {Mandolesi}, {Mann}, {Maris}, {Martin},
  {Martinez-Gonzalez}, {Masi}, {Matarrese}, {Matthai}, {Mazzotta},
  {Melchiorri}, {Mendes}, {Mennella}, {Mitra}, {Miville-Deschenes}, {Moneti},
  {Montier}, {Morgante}, {Mortlock}, {Munshi}, {Murphy}, {Naselsky}, {Natoli},
  {Netterfield}, {Norgaard-Nielsen}, {Novikov}, {Novikov}, {O'Dwyer}, {Oliver},
  {Osborne}, {Pajot}, {Pasian}, {Patanchon}, {Perdereau}, {Perotto},
  {Perrotta}, {Piacentini}, {Piat}, {Pinheiro Goncalves}, {Plaszczynski},
  {Pointecouteau}, {Polenta}, {Ponthieu}, {Poutanen}, {Prezeau}, {Prunet},
  {Puget}, {Rachen}, {Reach}, {Reinecke}, {Remazeilles}, {Renault},
  {Ricciardi}, {Riller}, {Ristorcelli}, {Rocha}, {Rosset}, {Rowan-Robinson},
  {Rubino-Martin}, {Rusholme}, {Sandri}, {Santos}, {Savini}, {Scott},
  {Seiffert}, {Shellard}, {Smoot}, {Starck}, {Stivoli}, {Stolyarov}, {Stompor},
  {Sudiwala}, {Sunyaev}, {Sygnet}, {Tauber}, {Terenzi}, {Toffolatti}, {Tomasi},
  {Torre}, {Tristram}, {Tuovinen}, {Umana}, {Valenziano}, {Vielva}, {Villa},
  {Vittorio}, {Wade}, {Wandelt}, {White}, {Yvon}, {Zacchei}, \&
  {Zonca}}]{2011arXiv1101.2028P}
{Planck Collaboration} {et~al.}, 2011, ArXiv e-prints

\bibitem[{{Remazeilles}, {Delabrouille} \& {Cardoso}(2011){Remazeilles},
  {Delabrouille}, \& {Cardoso}}]{2011MNRAS.410.2481R}
{Remazeilles} M., {Delabrouille} J., {Cardoso} J., 2011, \mnras, 410, 2481

\bibitem[{{Tegmark}(1998)}]{1998ApJ...502....1T}
{Tegmark} M., 1998, \apj, 502, 1

\bibitem[{{Tegmark} \& {Efstathiou}(1996)}]{1996MNRAS.281.1297T}
{Tegmark} M., {Efstathiou} G., 1996, \mnras, 281, 1297

\end{thebibliography}

\label{lastpage}

\end{document}